\documentclass[%
 reprint,
superscriptaddress,
amsmath,amssymb,
 aps,
 prl,
floatfix,
]{revtex4-2}

\usepackage{physics}
\usepackage{color}
\usepackage{xcolor}
\usepackage{graphicx}
\usepackage{dcolumn}
\usepackage{bm}
\usepackage{calc}
\usepackage{verbatim}
\usepackage{float}

\usepackage[mode=buildnew]{standalone}
\begin{document}

\title{Quantum optics of harmonic generation in the strongly driven Jaynes–Cummings-type system}

\author{M Suman}
 \email{suman.mohan@mbi-berlin.de}
\affiliation{Max Born Institute, Max-Born Stra{\ss}e 2A, 12489 Berlin, Germany}
\affiliation{Institute of Physics, Humboldt University Berlin, 12489 Berlin, Germany}

\author{Sili Yi}
 \email{Sili.Yi@mbi-berlin.de}
\affiliation{Max Born Institute, Max-Born Stra{\ss}e 2A, 12489 Berlin, Germany}
\affiliation{Institute of Physics, Humboldt University Berlin, 12489 Berlin, Germany}

\author{Maria Chekhova}
\affiliation{Max Planck Institute for the Science of Light, Staudtstr. 2, 91058 Erlangen, Germany}
\affiliation{Department of Physics, Friedrich-Alexander-Universit\"at Erlangen-N\"urnberg, Staudtstr. 7,
91058 Erlangen}
\affiliation{Technion – Israel Institute of Technology, 3200003 Haifa, Israel}

\author{Misha Ivanov}
\affiliation{Max Born Institute, Max-Born Stra{\ss}e 2A, 12489 Berlin, Germany}
\affiliation{Institute of Physics, Humboldt University Berlin, 12489 Berlin, Germany}
\affiliation{Technion – Israel Institute of Technology, 3200003 Haifa, Israel}

\begin{abstract}
We adapt the Jaynes-Cummings model to study the interface of cavity quantum electrodynamics with strong field and attosecond physics.
We show how multi-photon resonances in the Jaynes-Cummings system driven by a strong low-frequency classical light field lead to the generation of highly non-classical, quantum-correlated harmonics of the classical driver. Our treatment assumes no approximations, apart from the typical Jaynes-Cummings model assumption of only a few discrete quantum modes of light.  The paper is dedicated to Joseph Henry Eberly, whose remarkable research has left indelible mark on both strong field physics and quantum optics.  
\end{abstract}

\date{\today}

\maketitle

The Jaynes–Cummings (JC) model \cite{jaynes1963comparison} is one of the foundational stones of quantum optics and cavity quantum electrodynamics (QED) \cite{larson2021jaynes,shore1993jaynes}. In its original formulation, the JC model describes a nearly resonant interaction between a two-level atom and a single quantized light field cavity mode. 
Quoting from the 1979 paper by J.H. Eberly, N. B. Narozhny and 
J. J. Sanchez-Mondragon \cite{eberly1980periodic}, the JC model "... is the simplest fully quantized model of interest in NMR, quantum optics, quantum electronics, and resonance physics in general".

Seemingly on the opposite side of the AMO (atomic, molecular and optical) physics lies high harmonic generation, which results from a generally non-resonant interaction of intense laser pulses, often carrying in excess of $10^{13}$ photons, with matter. Given the very large number of both the incident and the generated (often $10^{6-7}$) photons, for the three decades since their discovery \cite{lewenstein1994theory,mcpherson1987studies,ferray1988multiple,lhuillier1993high,corkum1993plasma,kulander1993dynamics,krause1992high,schafer1993above}
the quantum optical properties of harmonic generation have been paid only scarce attention. 
This situation has changed recently, with the rapidly growing number of publications resembling the number of photons generated in an inverted medium \cite{lewenstein2021generation,yi2025generation, delapena2025fully,theidel2024evidence, theidel2026sub}.  Where only classical fields were originally and almost universally expected, squeezing and Schrödinger cat-like states for individual colors \cite{lewenstein2021generation,yi2025generation}  and non-classical correlations between different harmonic orders are now not only theoretically predicted \cite{delapena2025fully,yi2025generation}, but also experimentally observed \cite{theidel2024evidence, theidel2026sub}. 

However, the physical origin of these non-classical correlations, their apparent sensitivity to the intensity of the driving field and the specifics of the material system \cite{delapena2025fully,theidel2024evidence,theidel2026sub}, the conditions under which individual harmonic colors become squeezed or acquire Schrödinger cat-like features, 
and the possible link between non-classical features of a given harmonic and non-classical correlation between different harmonics remain unclear. 
The relative simplicity of the JC model explored here allows us to
firmly trace the origin of these effects to light-induced 
multi-photon resonances and cascade wave-mixing processes.
Reminiscent of shadows in Plato's "Allegory of the Cave" 
\cite{plato2021},  we find that the emergence of quantum correlations
between harmonics is a shadow of the nonclassical state produced for at least one of them.

\begin{figure}
    \centering
    \includegraphics[width=\linewidth]{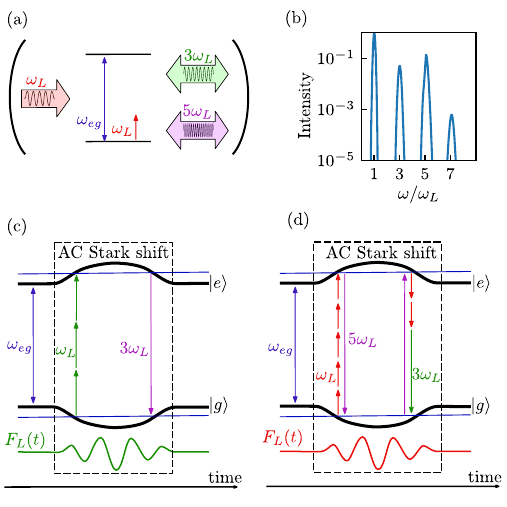}
    \caption{Harmonic generation in a strongly driven JC model. (a) Sketch of the setup, including a driven two-level system in a cavity tuned to
    the third and the fifth harmonic. 
    (b) Classical harmonic spectrum in the case of a five-photon resonance between the atom and the driving field. 
    (c) Dynamic Stark shift following the pulse envelope drives the 
    states into a multi-photon resonance (here, three-photon). 
    (d) Generated fifth harmonic (H5) can assist the generation of the third (H3), leading to quantum correlations between them if H5 is generated in a non-classical state, see text. 
    }
    \label{fig:FigureSetup}
\end{figure}

The setup and the physics are sketched in Fig.\ref{fig:FigureSetup}. 
Panel (a) shows a two-level atom inside a cavity. In contrast to the standard JC model setup, the system is driven by a strong classical light field at the 
frequency $\omega_L$, which is significantly smaller than the transition frequency $\omega_{eg}$, $\omega_L\ll \omega_{eg}$. Several harmonics of the classical driving
field are generated (Fig.\ref{fig:FigureSetup}a,b). The classical field strength $F_{L}$ is chosen such that predominantly harmonics H3 and H5
are generated ($\Omega_L\equiv d_{eg}F_L=0.2\omega_{eg}$ in Fig.\ref{fig:FigureSetup}b). 

In the absence of coupling between the quantum field fluctuations and the atom, the harmonics $N$ are produced in  
coherent states $|\alpha_{N}\rangle$, where (up to a phase)
\begin{equation}
\alpha_N= \mathcal{F}_N \int dt \  d(t) e^{iN\omega_L t}
    \label{eq:Alpha_N}
\end{equation}
and $\mathcal{F}_N$ is the vacuum field at the frequency $\omega_N=N\omega_L$. In the calculations below $\mathcal{F}_N$ are chosen such that $\alpha_N\sim 1$ for $N=3,5$. Other harmonics remain essentially in the vacuum states.

Let the cavity support only one of the two harmonics, e.g. $N=3$, which in this case is also nearly resonant with the atom (see panel c).  
As the low-frequency classical field turns on, 
the field-free states turn into the Floquet states, with their quasienergies approximately following the non-resonant Stark shift, see panel (c).
As the intensity changes during the pulse, the two Floquet states
come into a multi-photon resonance with the driver. 
In this case, the resonant harmonic 
($N=3$ in panel (c)) should acquire highly non-classical features
\cite{yi2025generation}. We will see below how these features depend on the detuning of the 
multi-photon transition between the Stark-shifted
states, and find that the emergence of the non-classical light is linked to the multi-photon resonance
in the laser-dressed system.

Second, we consider the case when two harmonics, $N=3$ and $N=5$ (referred to as H3 and H5 below), are supported by the cavity. One of them,  H5 in panel (d), is  also nearly resonant with the laser-driven atom. 
Now, the deviation of the resonant harmonic from the coherent state triggers non-classical correlations between the harmonics. 
One possible pathway towards such correlations 
is shown in panel (d).
Here, the resonant harmonic, H5 in this case, induces excitation to the Stark-shifted excited state, which 
leads to harmonic's nonclassical properties.
Now the third harmonic
can be resonantly generated, assisted by  
the two photons
of the strong classical driver.

Note that such pathway does not lead to quantum 
correlations if the harmonics are initially generated in  the coherent states, but it does lead to 
such correlations if at least one of the harmonics is generated in a non-classical state. 
Indeed, the short-time propagator associated with
the channel in panel (d) is $\hat U\simeq\hat 1-i\lambda \hat{a}^{\dagger}_3 \hat{a}_5$, where $\lambda$ is 
related to the coupling strengths, the classical driver, and the propagation time-interval. Let this propagator act on what is, initially, a product state
of H3 and H5, $|\chi\rangle=|\alpha_3\rangle|\chi^{(5)}\rangle$, 
but let H5 be in a 
superposition of two coherent states: $|\chi^{(5)}\rangle=c|\alpha_5\rangle+\tilde{c}|\tilde\alpha_5\rangle$. The result is
\begin{eqnarray} 
|\chi'\rangle\equiv \hat U|\chi\rangle &=&|\alpha_3\rangle|\chi^{(5)}\rangle
-i
\lambda \hat{a}^{\dagger}_3|\alpha_3\rangle|\tilde\chi^{(5)}\rangle
\nonumber
\\
|\tilde\chi^{(5)}\rangle &=& c\alpha_5|\alpha_5\rangle+\tilde c\tilde\alpha_5|\tilde\alpha_5\rangle
    \label{eq:UonChi}
\end{eqnarray}
Since both $|\alpha_3\rangle$ and $\hat{a}^{\dagger}_3|\alpha_3\rangle$, and  $|\chi^{(5)}\rangle$ and $|\tilde\chi^{(5)}\rangle$ are in general
linearly independent, the resulting wavefunction $|\chi'\rangle$ can no longer be factorized and the two harmonics become entangled. 

Since the emergence of the initial non-classical state of H5 requires resonance, and the resonance  has to be achieved between the Stark-shifted states, naturally the result for the 
developed correlations is sensitive to the intensity of 
the driving field, 
the  system-specific transition frequency, and to the frequency of the driving laser.

We now move on to the calculation. 
The Hamiltonian $\hat H_A$ of the two-level system, with eigenvalues $E_g$ and $E_e$ and eigenvectors $|g\rangle$ and $|e\rangle$, respectively, also includes the
interaction with the classical time dependent electric field $F_{L}(t)=F_{L}f(t)\cos{(\omega_Lt)}$ ($f(t)$ is the pulse envelope)
\begin{equation}
    \hat H_{A}(t) = \hat H_0 -\frac{\Omega_{L}f(t)}{2}
    \left( e^{i\omega_L t} + e^{-i\omega_L t}\right)\hat S 
    \label{eq:ClassicalHamiltonian}
\end{equation}
where $\Omega_{L}=d_{eg}F_{cl}$ is the Rabi coupling for the classical field, $d_{eg}$ is the transition matrix element,  and 
$\hat S=\hat\sigma_{+}+\hat\sigma_{-}$ describes the creation and annihilation of excitations in the atom.
Interaction with the quantized field, in the interaction picture for the quantum fields, is
\begin{align}
    \hat H_{int} & = \hat{S}
    \left[\frac{\Omega_3}{2}\hat A_3(t)+ \frac{\Omega_5}{2}\hat A_5(t)\right]
    \\
    \hat A_N(t)& =(\hat a^\dagger_N e^{iN\omega_L t} + \hat a_N e^{-iN\omega_L t})
     \label{eq:QuantumH_int}
\end{align}
Here $\Omega_{N}=\mathcal{F_N}d_{eg}$ are the vacuum Rabi couplings for harmonics $N=3,5$,  $\mathcal{F_N}$ is the vacuum field for the harmonic $N$. Finally,  $\hat a^\dagger_N$ and $\hat a_N$ are the creation and annihilation operators for the harmonic $N$.  The
initial state is 
$|\Psi_{in}\rangle = |g\rangle|\chi_{in}\rangle$, 
where $|\chi_{in}\rangle=|0_3,0_5\rangle$ is the vacuum state
of both harmonics. For the numerical simulations, we set $\Omega_3 = 0.0041\sqrt{\omega_L}$ a.u. and $\Omega_5 = 0.0031\sqrt{\omega_L}$ ($\Omega_3 \simeq 0.0098\omega_{eg}$ and 
$\Omega_5 \simeq 0.0059\omega_{eg}$ for $\omega_L=\omega_{eg}/5$.)

First, we introduce exact dressed states 
$|\phi_{g,e}(t)\rangle$ corresponding to the solutions of the semi-classical 
time-dependent Schrodinger equation with the semi-classical Hamiltonian $\hat H_A$ Eq.(\ref{eq:ClassicalHamiltonian}), starting in the field-free ground $|g\rangle$ and excited
$|e\rangle$ states correspondingly.  Just like the field-free states $|g,e\rangle$, the dressed states $|\phi_{g,e}(t)\rangle$ are orthogonal and form a complete basis 
for the material system. 
The time evolution of the full "atom + quantum light"  wavefunction can then be expanded in this basis,
\begin{equation}
    |\Psi(t)\rangle = |\phi_g(t)\rangle|\chi_g (t)\rangle + |\phi_e(t)\rangle|\chi_e (t)\rangle
    \label{eq:Wavefunction1}
\end{equation}
with quantum states of light $|\chi_g(t)\rangle$, $|\chi_e(t)\rangle$ correlated to $|\phi_g(t)\rangle$, $|\phi_e(t)\rangle$. Equations for $|\chi_{e,g} (t)\rangle$ are \cite{yi2025generation} 
\begin{equation}
\begin{split}
    i \frac{\partial}{\partial t} |\chi_g (t)\rangle &= \mathbf{\Omega}.\hat{\mathbf{A}}(t) \left[S_{gg}(t) |\chi_g(t)\rangle +S_{ge}(t) |\chi_e(t)\rangle\right]
    \\
    i \frac{\partial}{\partial t} |\chi_e (t)\rangle &= \mathbf{\Omega}.\hat{\mathbf{A}}(t) \left[S_{eg}(t) |\chi_g(t)\rangle +S_{ee}(t) |\chi_e(t)\rangle\right]  
\end{split}
\label{eq:SchrodingerEquation}
\end{equation}
where     
$\mathbf{\Omega}.\hat{\mathbf{A}}(t)$ is the shorthand  for $\sum_{N=3,5}\Omega_{N} \hat A_N(t)$ and 
\begin{equation}
    S_{ij}(t)= \langle \phi_i(t)|\hat S|\phi_j(t)\rangle
\end{equation}
are the dipoles, normalized to $d_{eg}$, induced
in each of the dressed states (for $i=j$) and 
the transition dipoles between them (for $i\neq j$).

We set the transition energy $\omega_{eg}=0.057$ a.u., similar to that between the ground and the first excited states in Rb atom, corresponding to the wavelength $\lambda\simeq 800$ nm.  We then 
solve the semiclassical TDSE without the quantum field, 
for different wavelengths $\lambda_L$ of the driver, starting first in the field-free atomic 
ground and then in the field-free excited state. This yields the dressed states
$|\phi_{g,e}(t)\rangle$. 
The laser pulse is $F_L(t)=F_Lf(t)\cos\omega_Lt$, with
tunable $\omega_L$, $\Omega_L=d_{eg}F_L = 0.2\omega_{eg} \simeq 0.0117$ a.u., and the laser envelope is $f(t)=e^{-t^2 /(2\tau_0^2)}$, $\tau_0 = 100$ fs. 

\begin{figure}
    \centering
    \includegraphics[width=\linewidth]{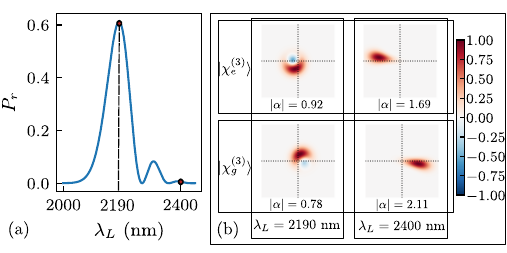}
    \caption{Wigner negativity near resonance. (a) Excitation probability $P_r = |\langle\phi_g(t)|e\rangle|^2$ vs driving wavelength. (b) Wigner function of light state correlated to dressed excited state $|\chi_e^{(3)}\rangle$ and dressed ground state $|\chi_g^{(3)}\rangle$ for $\lambda_L = 2190$ nm (induced resonance) and $\lambda_L = 2400$ nm (field free resonance). Even though the Wigner negativity is absent in this regime, significant squeezing is observed.}
    \label{fig:FigureResonances}
\end{figure}

Starting in the ground state $\ket{g}$ and projecting 
$|\phi_g(t)\rangle$ after the end of the 
laser pulse on the field-free state
$\ket{e}$ yields the final excitation probability
into the state $\ket{e}$.  
Representative results are shown 
in Figure \ref{fig:FigureResonances} (a), for
the case of a three-photon resonance. 
The probability 
shows the characteristic oscillations, originally described by Ernst Carl Gerlach St{\"u}ckelberg for collisions \cite{stuckelberg1932theory}, 
but also appearing in Stark-shift-induced multi-photon resonances \cite{Bengs22}. These oscillations are
associated with the interference of the two 
resonant excitations at the rising and falling edges 
of the pulse. The Stark shift 
of the transition energy is clearly visible, as the 
maximum excitation is achieved for $\lambda_L\simeq 2190$ nm, not 
for the field-free resonance $\lambda_L\simeq 2400$ nm.

Next, we include coupling to the quantum field and solve
Eqs.(\ref{eq:SchrodingerEquation}). 
 The wavefunctions 
$|\chi_{g,e}(t)\rangle$ are expanded into the complete basis of the Fock states $|n_N\rangle$ for both harmonics $N=3,5$, 
$|\chi_{g,e}(t)\rangle=\sum_{n_3,n_5}C_{e,g}(n_3,n_5;t)|n_5\rangle|n_3\rangle$. The coupled equations for the corresponding amplitudes $C_{e,g}(n_3,n_5;t)$ are solved numerically exactly, without using the rotating wave approximation.

Setting $S_{eg}=S_{ge}=0$ in Eqs.(\ref{eq:SchrodingerEquation}) neglects the effect of the generated harmonics on the material system. The generated light states are then always coherent states, regardless of how complex $S_{gg}(t)$ or $S_{ee}(t)$ are. 

Including $S_{eg}, S_{ge}$ in the equations but neglecting the coupling between the two harmonics implements the decoupling ansatz \cite{lange2023electron} and yields
\begin{equation}
\begin{split}
    i \frac{\partial |\chi^{(N)}_g\rangle}{\partial t}  =\Omega_{N} \hat A_N(t) \left[S_{gg}(t) |\chi^{(N)}_g\rangle +S_{ge}(t) |\chi^{(N)}_e\rangle\right]
    \\
    i \frac{\partial |\chi^{(N)}_e\rangle }{\partial t} = \Omega_{N} \hat A_N(t) \left[S_{eg}(t) |\chi^{(N)}_g\rangle +S_{ee}(t) |\chi^{(N)}_e\rangle\right]  
\end{split}
\label{eq:SEDecoupling}
\end{equation}
for the states of individual harmonics $|\chi^{(N)}_{g,e}(t)\rangle$ correlated to the dressed ground and excited states. 

Figure \ref{fig:FigureResonances} shows the role of
three-photon resonance for the quantum state of H3. The calculations are done using the 
decoupling ansatz. The Wigner functions  for $|\chi^{(3)}_{g,e}(t)\rangle$, 
for the two different 
values of the driving field 
wavelength $\lambda_L$ are shown in the figure \ref{fig:FigureResonances} (b)

We now include coupling between harmonics in
Eqs.(\ref{eq:SchrodingerEquation}). To assess the origin of entanglement, 
we re-write Eq.\eqref{eq:SchrodingerEquation} for $|\chi_g (t)\rangle$ as 
\begin{eqnarray}  
    &&i \frac{\partial}{\partial t} |\chi_g (t)\rangle = \mathbf{\Omega}.\hat{\mathbf{A}}(t) S_{gg}(t) |\chi_g(t)\rangle -i\mathbf{\Omega}.\hat{\mathbf{A}}(t)S_{ge}(t)\times
      \nonumber
    \\
    &&\times\int^t \!\!\! dt'\mathbf{\Omega}.\hat{\mathbf{A}}(t') \left[ S_{eg}(t') |\chi_g(t')\rangle +S_{ee}(t') |\chi_e(t')\rangle \right]
\label{eq:SecondOrder}
\end{eqnarray}
The term quadratic in $\mathbf{\Omega}.\hat{\mathbf{A}}(t)$ includes $\hat a_3^\dagger \hat a_5^\dagger$, $\hat a_3^\dagger \hat a_5$, $\hat a_3 \hat a_5^\dagger$, and $\hat a_3 \hat a_5$. Thanks to resonances created by the "support photons" $\omega_L$ available from the strong driving field, absorption of one, e.g. H5, can be accompanied by the emission of the other, e.g. H3, augmented by the two $\omega_L$, see Fig.\ref{fig:FigureSetup}(d). The converse, where
the absorption of H3 augmented by two $\omega_L$ is accompanied by the emission of H5, is also possible. 
The corresponding operator sequences, 
$\hat a_3^\dagger \hat a_5$ and $\hat a_3 \hat a_5^\dagger$, do not individually generate entanglement when applied to coherent states. However, if $|\chi_{e,g}(t')\rangle$ are not  coherent states -- which occurs thanks to resonant excitation \cite{yi2025generation}, entanglement 
develops naturally, see Eq.\ref{eq:UonChi}. Even if e.g. $|\chi_{g}\rangle=|\alpha_3\rangle |\alpha_5\rangle$ with 
nonzero $\alpha_{3,5}$, the action of  
$\hat a_3^\dagger \hat a_5+\hat a_3 \hat a_5^\dagger$ 
generates entanglement as $\hat a_3^\dagger|\alpha_3\rangle$
and $\hat a_5^\dagger|\alpha_5\rangle$ are no longer coherent states.

The Cauchy-Schwarz inequality, which holds for classical observables, can be violated in the presence of quantum correlations \cite{VoilationCSInequality01}. We therefore compute the parameter $R$ defined as
\begin{equation}
    R_{35} = 
    \frac{\langle a_3^\dagger a_5^\dagger a_5 a_3 \rangle^2}{\langle a_3^{\dagger^2}a_3^2 \rangle \langle a_5^{\dagger^2}a_5^2 \rangle}
\end{equation}
Here $\langle a_i^\dagger a_j^\dagger a_i a_j \rangle =g_{ij}^{(2)}(0)$ for $i,j=3,5$ are the second-order intensity-intensity correlation functions. The Cauchy-Schwartz inequality is violated when $R_{35} > 1$. 

\begin{figure}
    \centering
    \includegraphics[width=1.0\linewidth]{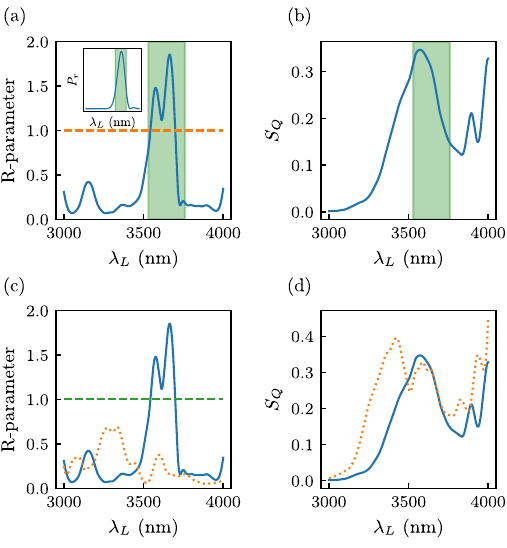}
    \caption{Correlation between harmonics for five-photon resonance. (a) Violation of the Cauchy-Schwartz inequality $R > 1$ for H3 and H5, vs the driving field wavelength. The inset shows excitation probability $P_r = |\langle\phi_g(t)|e\rangle|^2$ vs the driving field wavelength in the case of a five-photon resonance, for the classical driver only. 
    The green-shaded region marks the effective width of the resonance. (b) Effective entanglement entropy $S_Q$ between H3 and H5 vs the driving field wavelength. (c,d) Comparison of $R$ (c) and $S_Q$ (d) for two values of $F_L$. Blue solid line and orange dotted line represents, $F_L = 0.038$ V/\AA and $F_L = 0.046$ V/\AA, respectively.}
    \label{fig:Figure_R}
\end{figure}
Figure \ref{fig:Figure_R} (a) shows $R_{35}$ vs the driving field wavelength; we see that $R$ exceeds unity near the field-induced multi-photon resonance. 
Panel (b) shows effective entanglement entropy between the two harmonics.
The entanglement entropy characterizes correlations in a bi-partite system and is equal to zero for product states. Here we have two bi-partite wavefunctions $|\chi_{e,g}\rangle$, describing both H3 and H5, and correlated to the two atomic states. We therefore  define $S_Q = S_R^{(e)} W_e + S_R^{(g)} W_g$, where $S_R^{(e,g)}$ is the standard entanglement entropy for $|\chi_{e,g}\rangle$, while $W_{e,g}$ are the populations of the two states (given by the norms of $|\chi_{e,g}\rangle$.) We see that $S_Q$ in panel (b) is non-zero even when $R_{35}<1$, but it also develops maxima near the wavelengths where $R_{35} > 1$ and 
where the system undergoes five-photon resonance, as shown with green shading. 
Thus, significant entanglement between H3 and H5 is generated near multi-photon resonances, where the 
individual harmonics are strongly non-classical. 

The entanglement entropy remains non-zero even outside the optimal resonant
condition, showing that $R$ underestimates the degree of entanglement. 

The $R$ parameter is also sensitive to the driving field strength, and
appears to decrease in stronger fields as previously noted in more complex systems, see e.g.\cite{theidel2024evidence,delapena2025fully}. Again, we see that (i)  $R<1$ does not imply the absence of entanglement between the harmonics, and that (ii) resonances
between dressed states correspond to the highest
correlation between the harmonics. 

In conclusion, we have discussed the extension of the JC model, where a two-level atom in a cavity is driven by a strong classical field and the cavity supports several harmonics of the driver. 
The relative simplicity of the JC model allows one to efficiently explore the interplay of the many parameters that control the quantum nonlinear-optical response of matter, the emergence of non-classical states of harmonic light, and the development of light-matter and light-light entanglement.
While multi-photon resonances induced between the Stark-shifted states do lead to the violation of the Cauchy-Schwartz inequality for the generated harmonics, the lack of such violation does not necessarily imply the lack of correlations between them, as shown by $S_Q$.

We acknowledge support of ANR-DFG project "Generation of bright non-classical light based on high harmonics and its use in quantum spectroscopy,” project No. 545591821, and ISF-DFG project "Quantum optics of high harmonic generation in resonant media", project No. 560535838. S.M. acknowledges support from the European Union’s Horizon Europe research and innovation programme under the Marie Skłodowska-Curie grant agreement No 101168628 (project Qu-ATTO). S.Y. acknowledges the support of the CSF No. 202308080044. 

\bibliography{refs}

\end{document}